\documentclass{actams-en}
\usepackage{harpoon}
\usepackage{threeparttable}
\usepackage{tabularx,multirow,bigdelim}
\usepackage{graphics,graphicx,subfigure}
\usepackage{booktabs}
\usepackage{tabularx}
\usepackage{array}
\usepackage{float}
\usepackage{placeins}
\usepackage{bm}
\numberwithin{equation}{section} 

\begin{document}

\Volume{2026}{}{44}{1}
\DOI{xx}
\PageNum{1}

\EditorNote{This work was supported by China's National Natural Science Foundation Grant Nos.~12575033, 12575035, 12304257 and 12322501.
Y.T. acknowledges support from the Natural Science Foundation of Sichuan Province Grant No.~2026NSFSCZY0124.
Computing resources were provided by the Interdisciplinary Intelligence Supercomputing Center of Beijing Normal University, Zhuhai.}

\EditorNote{$^*$Corresponding author: Zengru Di; 
E-mails: zdi@bnu.edu.cn}

\abovedisplayskip 6pt plus 2pt minus 2pt
\belowdisplayskip 6pt plus 2pt minus 2pt

\def\no{\nonumber}
\def\R{{\Bbb R}}
\newenvironment{prof}[1][Proof]{\indent\textbf{#1}\quad }
{\hfill $\Box$\vspace{0.7mm}}
\def\q{\quad}
\def\qq{\qquad}
\allowdisplaybreaks[4]

\AuthorMark{F. Wang, W. Liu, K. Qi, Y. Tang \,\&\, Z. Di}

\TitleMark{\uppercase{Third-Order MIPA-type Transitions}}

\title{\uppercase{Response Morphologies of a Canonical Fluctuation Diagnostic Across Ehrenfest Phase Transitions}}

\author{
\sl{Fangfang Wang}$^{1,4,5}$\quad
\sl{Wei Liu}$^{2}$\quad
\sl{Ying Tang}$^{3}$\quad
\sl{Zengru Di}$^{1,4,5,*}$
}
{
$^{1}$Department of Systems Science, Faculty of Arts and Sciences, Beijing Normal University, Zhuhai 519087, China\\
$^{2}$College of Science, Xi'an University of Science and Technology, Xi'an 710600, China\\
$^{3}$Institute of Fundamental and Frontier Sciences, University of Electronic Science and Technology of China, Chengdu 611731, China\\
$^{4}$International Academic Center of Complex Systems, Beijing Normal University, Zhuhai 519087, China\\
$^{5}$School of Systems Science, Beijing Normal University, Beijing 100875, China\\
E-mail: zdi@bnu.edu.cn
}

\maketitle

\begin{abstract}
Microcanonical inflection-point analysis identifies MIPA-type higher-order transition structures from derivatives of the microcanonical entropy. Until recently, however, there was no canonical formulation for probing the corresponding fluctuation-level behavior directly from measurable energy fluctuations without reconstructing the density of states. Previous work addressed this limitation by introducing a canonical fluctuation diagnostic that quantifies energy-fluctuation asymmetry. Here, we investigate how this diagnostic behaves across representative first-, second-, and third-order phase transitions in the Ehrenfest classification. We consider the eight-state Potts model, the two-dimensional Ising model, and ideal three-dimensional Bose--Einstein condensation. The Potts and Ising systems exhibit a robust paired-extremum structure of the diagnostic near their respective transition regions, whereas ideal Bose--Einstein condensation exhibits a discontinuous response at the condensation temperature in the thermodynamic limit. These results show that a single fluctuation-based observable can display qualitatively distinct response morphologies across thermodynamic regimes, reflecting sensitivity to the nature of the underlying phase transition rather than serving as a direct classifier of Ehrenfest order.
\end{abstract}


\MRSubClass{82B20}

\section{Introduction}

Phase transitions are conventionally characterized by nonanalyticities in thermodynamic potentials or their derivatives~\cite{StatisticalMechanics2001,Landautheory1999}. 
Within the Ehrenfest classification~\cite{Ehrenfest1933}, first-order phase transitions exhibit discontinuities in the first derivative of the free energy, whereas second-order phase transitions are associated with singularities in the second derivative. 
Third-order phase transitions, in contrast, are defined by continuity of the free energy and its first two derivatives, with nonanalytic behavior appearing only in the third derivative. 
While this classification provides a systematic thermodynamic description based on singularities of the free energy and their scaling behavior near criticality~\cite{FisherScalingTheory1972,Wilson1983RenormalizationGroup}, it does not explicitly address the organization of additional fluctuation-level structures near phase transitions.
In this work, ``phase transition'' specifically denotes a thermodynamic transition defined by a free-energy nonanalyticity, whereas MIPA-type higher-order transitions refer to fluctuation-level structures that may appear near lower-order transitions and should not, in general, be regarded as Ehrenfest phase transitions.

In parallel to this conventional classification, a distinct line of research has revealed that systems near phase transitions may exhibit fluctuation-level structures reflecting higher-order organization of thermodynamic fluctuations~\cite{Gross2001BOOK}. 
These structures are not part of the Ehrenfest classification itself. 
They can appear in finite systems and, in some cases such as the two-dimensional Ising model, persist in the thermodynamic limit as robust signatures of fluctuation asymmetry and structural reorganization. 
In particular, microcanonical entropy inflection points provide a systematic route for identifying finite-system transition signatures~\cite{Microcanonicalentropyinflectionpoints2011}, and microcanonical inflection-point analysis (MIPA) further classifies such structures through derivatives of the microcanonical entropy, revealing MIPA-type independent and dependent third-order transitions in a variety of classical models including Ising and Potts-type systems~\cite{QiClassification2018,wangff}.

Recent developments have also connected microcanonical inflection-point analysis with parametric-curve representations and partition-function zeros, further clarifying its relation to finite-system transition diagnostics~\cite{Rocha2025MIPAParametric}.
In addition, a canonical formulation based on energy cumulants has been proposed, establishing an experimentally accessible diagnostic that reproduces leading MIPA-type behavior in the single-saddle regime and extends its applicability beyond explicit density-of-states reconstruction~\cite{Wang2026Canonical}. 
Complementary studies based on geometric observables and spectral decompositions further suggest that MIPA-type higher-order transitions correspond to distinct forms of cluster reorganization and mode redistribution, indicating that they encode physically meaningful fluctuation structures rather than numerical artifacts~\cite{Liu2026EigenMicrostate,Sitarachu2020ThirdOrderIsing,Sitarachu2022IsingThirdOrder,Liu2025PottsGeometry,Bachmann2025Histogram}. 
These studies have been extended beyond lattice spin systems to a range of classical models, further supporting the MIPA-type third order transitions across different microscopic realizations~\cite{BelHadjAissa2021KT_XY,DiCairano2024Topological,Liu2024MV,Qi2019FlexiblePolymers,SemiflexiblePolymers2023,Wang2026ThirdOrderWS}.

Despite these advances, most existing studies focus on identifying MIPA-type structures within individual models or within specific phase-transition classes. 
As a result, a systematic understanding of how a fixed fluctuation-based diagnostic behaves across different Ehrenfest classes of primary phase transitions remains incomplete. 
In particular, it remains unclear whether the morphological patterns associated with MIPA-type third-order behavior represent a universal response of fluctuation asymmetry, or whether they depend qualitatively on the nature of the underlying thermodynamic nonanalyticity.

These considerations motivate the present work. 
We study how a fixed canonical third-order fluctuation diagnostic responds when applied to systems belonging to different Ehrenfest classes of phase transitions, spanning first-order, second-order, and third-order phase transitions. 
Rather than introducing a new diagnostic or redefining phase-transition classes, we focus on the response morphology of a single established observable across different thermodynamic regimes, in order to clarify which aspects of higher-order fluctuation structure are intrinsic to the diagnostic and which aspects depend on the underlying phase transition. 
To this end, we consider three representative systems: the eight-state Potts model, the two-dimensional Ising model, and ideal three-dimensional Bose--Einstein condensation, serving as prototypical examples of first-order, second-order, and third-order phase transitions in the Ehrenfest sense. 
We show that the same diagnostic exhibits qualitatively distinct response morphologies across these regimes, indicating that the morphology of $\Xi(T)$ encodes not only local fluctuation-asymmetry structure but also information about the nature of the primary thermodynamic nonanalyticity.

\section{Models and methods}
\label{sec:models_methods}

In this section, we introduce the representative models and the methods used to analyze the canonical third-order diagnostic. 
The models are chosen to cover three Ehrenfest classes of phase transitions: a first-order phase transition, a second-order phase transition, and a third-order phase
transition. 
We first describe how the diagnostic is evaluated or obtained for the three systems, and then give the thermodynamic derivation underlying the nonanalytic-jump morphology in ideal Bose--Einstein condensation.

\subsection{Models}
\label{subsec:models}

\subsubsection{Potts model}

The ferromagnetic eight-state Potts model is used as the representative system with a
first-order phase transition. 
On an $L\times L$ square lattice with periodic boundary conditions, the Hamiltonian is \cite{potts1,WuRMP1982}
\begin{equation}
    E
    =
    -J\sum_{\langle ij\rangle}\delta(\sigma_i,\sigma_j),
    \qquad
    \sigma_i\in\{0,1,\ldots,q-1\},
    \qquad
    q=8,
    \label{eq:potts_hamiltonian}
\end{equation}
where $\delta(\sigma_i,\sigma_j)$ is the Kronecker delta. 
For the square-lattice Potts model, the transition temperature is
\begin{equation}
    \frac{T_c}{J}
    =
    \frac{1}{\ln(1+\sqrt{q})}.
    \label{eq:potts_Tc}
\end{equation}
For $q=8$, this gives
\begin{equation}
    \frac{T_c}{J}
    \simeq
    0.745.
\end{equation}
Since the square-lattice Potts phase transition is first order for $q>4$, the eight-state
model provides a standard first-order benchmark with phase coexistence and latent heat
effects~\cite{WuRMP1982}. 
The MIPA-type third-order transitions around the coexistence region have been
identified in previous finite-size and canonical analyses~\cite{wangff,Wang2026Canonical}.

\subsubsection{Ising model}

The two-dimensional ferromagnetic Ising model is used as the representative system
with a second-order phase transition. 
On a square lattice, its Hamiltonian is \cite{ising}
\begin{equation}
    E
    =
    -J\sum_{\langle ij\rangle}s_i s_j,
    \qquad
    s_i=\pm 1,
    \qquad
    J>0,
    \label{eq:ising_hamiltonian}
\end{equation}
where $\langle ij\rangle$ denotes nearest-neighbor pairs counted once. 
We set $J=1$ unless otherwise stated. 
In the thermodynamic limit and at zero external field, the square-lattice Ising model
has a second-order phase transition at \cite{Onsager1944}
\begin{equation}
    T_c
    =
    \frac{2J}{\ln(1+\sqrt{2})}
    \simeq 2.269
    \quad
    (J=1).
    \label{eq:ising_Tc}
\end{equation}
This model provides a clean benchmark for studying how the canonical third-order
diagnostic behaves around a second-order phase transition. 
The MIPA-type third-order transitions around the Ising critical point are based on
previous microcanonical and canonical analyses~\cite{Sitarachu2022IsingThirdOrder,Wang2026Canonical}.

\subsubsection{Ideal three-dimensional Bose gas}
Ideal three-dimensional Bose--Einstein condensation is used as the representative system
with a third-order phase transition in the Ehrenfest classification~\cite{Ehrenfest1933}. 
In contrast to the Ising and Potts models, where MIPA-type third-order transitions appear
as fluctuation-level structures around lower-order phase transitions, Bose--Einstein
condensation represents a case in which the primary thermodynamic nonanalyticity itself
occurs in the third derivative of the free energy. 
For a noninteracting Bose gas with single-particle energies~\cite{Ziff1977IdealBoseGas,StatisticalMechanics2001}
\begin{equation}
    \epsilon_{\bm{k}}
    =
    \frac{\hbar^2 k^2}{2m},
    \label{eq:bec_energy}
\end{equation}
the condensation temperature is determined by the saturation of the excited-state
population. 
In the thermodynamic limit,
\begin{equation}
    T_c
    =
    \frac{2\pi\hbar^2}{m k_B}
    \left[
        \frac{N}{V\zeta(3/2)}
    \right]^{2/3}.
    \label{eq:bec_Tc}
\end{equation}
This system differs fundamentally from the Ising and Potts models. 
Here the primary phase transition itself is third order in the Ehrenfest classification,
rather than a lower-order phase transition accompanied by MIPA-type higher-order
transitions.

\subsection{Methods}
\label{subsec:methods}

\subsubsection{Numerical and reweighting procedures}

For all three models, we analyze the canonical third-order diagnostic \cite{Wang2026Canonical}
\begin{equation}
    \Xi(T)
    =
    \frac{\kappa_3(T)}{[\kappa_2(T)]^3},
    \label{eq:Xi_def}
\end{equation}
where $\kappa_2(T)$ and $\kappa_3(T)$ are the second and third cumulants of the total
energy. 
The purpose of the present work is not to rederive the full Ising or Potts results, but to
compare their canonical morphology with that of a third-order phase transition.

For the eight-state Potts model, the canonical diagnostic is evaluated from
density-of-states or canonical reweighting data \cite{parallel1996,Wang2001MultipleRangeRW,Berg1991MulticanonicalAlgorithms,Berg1992MulticanonicalEnsemble,Wang2001FlatHistogramPRE}. 
Given an estimate of the density of states $g(E)$, the canonical energy distribution at
temperature $T$ is
\begin{equation}
    P_T(E)
    =
    \frac{g(E)e^{-\beta E}}{Z(\beta)},
    \qquad
    \beta=\frac{1}{T},
    \label{eq:canonical_reweighting}
\end{equation}
where
\begin{equation}
    Z(\beta)=\sum_E g(E)e^{-\beta E}.
\end{equation}
The cumulants $\kappa_2(T)$ and $\kappa_3(T)$ are then computed from this canonical
energy distribution. 
This procedure locates the MIPA-type third-order transitions manifested as
paired signed extrema of $\Xi(T)$ around the first-order coexistence region.

For the two-dimensional Ising model, we use the thermodynamic-limit canonical result
obtained from Onsager's exact solution \cite{Onsager1944} and its associated energy cumulants. 
This provides a sampling-free reference for the second-order phase transition and for
the MIPA-type third-order transitions manifested as paired extrema of $\Xi(T)$ around
the critical point.

For ideal three-dimensional Bose--Einstein condensation, we compute the heat capacity
and the cumulant-ratio diagnostic directly from the thermodynamic-limit Bose-gas
relations. 
Temperatures are measured in units of $T_c$. 
For $T<T_c$, the fugacity is fixed at $z=1$, and the heat capacity per particle is
\begin{equation}
    \frac{C_V^-}{N k_B}
    =
    \frac{15}{4}
    \frac{\zeta(5/2)}{\zeta(3/2)}
    \left(
        \frac{T}{T_c}
    \right)^{3/2}.
    \label{eq:bec_CV_below}
\end{equation}
For $T>T_c$, the fugacity $z(T)$ is determined by
\begin{equation}
    g_{3/2}(z)
    =
    \zeta(3/2)
    \left(
        \frac{T_c}{T}
    \right)^{3/2},
    \label{eq:bec_fugacity}
\end{equation}
where $g_\nu(z)=\mathrm{Li}_\nu(z)$ is the Bose function. 
The heat capacity above the condensation point is
\begin{equation}
    \frac{C_V^+}{N k_B}
    =
    \frac{15}{4}
    \frac{g_{5/2}(z)}{g_{3/2}(z)}
    -
    \frac{9}{4}
    \frac{g_{3/2}(z)}{g_{1/2}(z)}.
    \label{eq:bec_CV_above}
\end{equation}
The derivative $dC_V/dT$ is evaluated on a fine temperature grid, and $\Xi(T)$ is then
computed from the canonical cumulant identities given below.

\subsubsection{Thermodynamic derivation for the BEC jump}

The key point of the ideal Bose gas is that its heat capacity is continuous at the
condensation temperature,
\begin{equation}
    C_V^-(T_c)=C_V^+(T_c),
    \label{eq:bec_CV_continuous}
\end{equation}
whereas its temperature derivative is discontinuous,
\begin{equation}
    \left.
    \frac{dC_V^-}{dT}
    \right|_{T_c}
    \neq
    \left.
    \frac{dC_V^+}{dT}
    \right|_{T_c}.
    \label{eq:bec_dCV_jump}
\end{equation}
Since
\begin{equation}
    C_V
    =
    -T\frac{\partial^2 F}{\partial T^2},
    \label{eq:CV_free_energy}
\end{equation}
we obtain
\begin{equation}
    \frac{\partial^3 F}{\partial T^3}
    =
    -\frac{1}{T}\frac{dC_V}{dT}
    +
    \frac{C_V}{T^2}.
    \label{eq:F3_CV}
\end{equation}
Therefore, the discontinuity of $dC_V/dT$ implies a discontinuity in the third
temperature derivative of the free energy. 
This establishes ideal three-dimensional Bose--Einstein condensation as a third-order
phase transition.

In the following cumulant identities, we use units with $k_B=1$, and $C_V$ denotes the total heat capacity.
The normalized heat capacity $C_V/(N k_B)$ is used only for plotting and for the Bose-gas formulas above.
The corresponding canonical energy cumulants satisfy
\begin{equation}
    \kappa_2(T)
    =
    T^2 C_V(T),
    \label{eq:kappa2_CV}
\end{equation}
and
\begin{equation}
    \kappa_3(T)
    =
    T^4\frac{dC_V}{dT}
    +
    2T^3 C_V(T).
    \label{eq:kappa3_CV}
\end{equation}
Thus,
\begin{equation}
    \Xi(T)
    =
    \frac{
        T^4\dfrac{dC_V}{dT}
        +
        2T^3C_V(T)
    }{
        [T^2C_V(T)]^3
    }.
    \label{eq:Xi_CV}
\end{equation}
Because $C_V$ is continuous but $dC_V/dT$ is discontinuous at $T_c$, the diagnostic
$\Xi(T)$ directly inherits this nonanalyticity and displays a jump at the transition
point. 
This is the expected canonical morphology of a third-order phase transition.

In contrast, the Ising and Potts models have lower-order phase transitions. 
For them, the paired signed extrema of $\Xi(T)$ are interpreted as MIPA-type
third-order transitions generated by fluctuation-asymmetry reorganization around the
primary phase transition, not as third-order phase transitions themselves.
For comparison, finite-$N$ curves are also shown in Figs.~\ref{fig:bec_cv} and~\ref{fig:bec_xi}. 
They are obtained from the finite-size Bose gas calculation using the same cumulant relations, and approach the thermodynamic-limit result as $N$ increases.

\section{Canonical morphologies of third-order behavior}
\label{sec:results_morphology}

Having introduced the diagnostic, we now compare its response morphology in the three representative systems. 
The comparison focuses on whether $\Xi(T)$ forms separated signed extrema around a lower-order transition or instead develops a direct nonanalytic jump at a genuine third-order transition.

\subsection{MIPA-type third-order structures around a first-order phase transition: Potts model}
\label{subsec:potts_results}

We first consider the square-lattice eight-state Potts model, which exhibits a first-order phase transition at $T_c/J=1/\ln(1+\sqrt{8}) \simeq 0.745$. In finite systems, this transition is rounded due to phase coexistence, leading to strong but smooth changes in the energy distribution.

\begin{figure}
    \centering
    \includegraphics[width=0.5\linewidth]{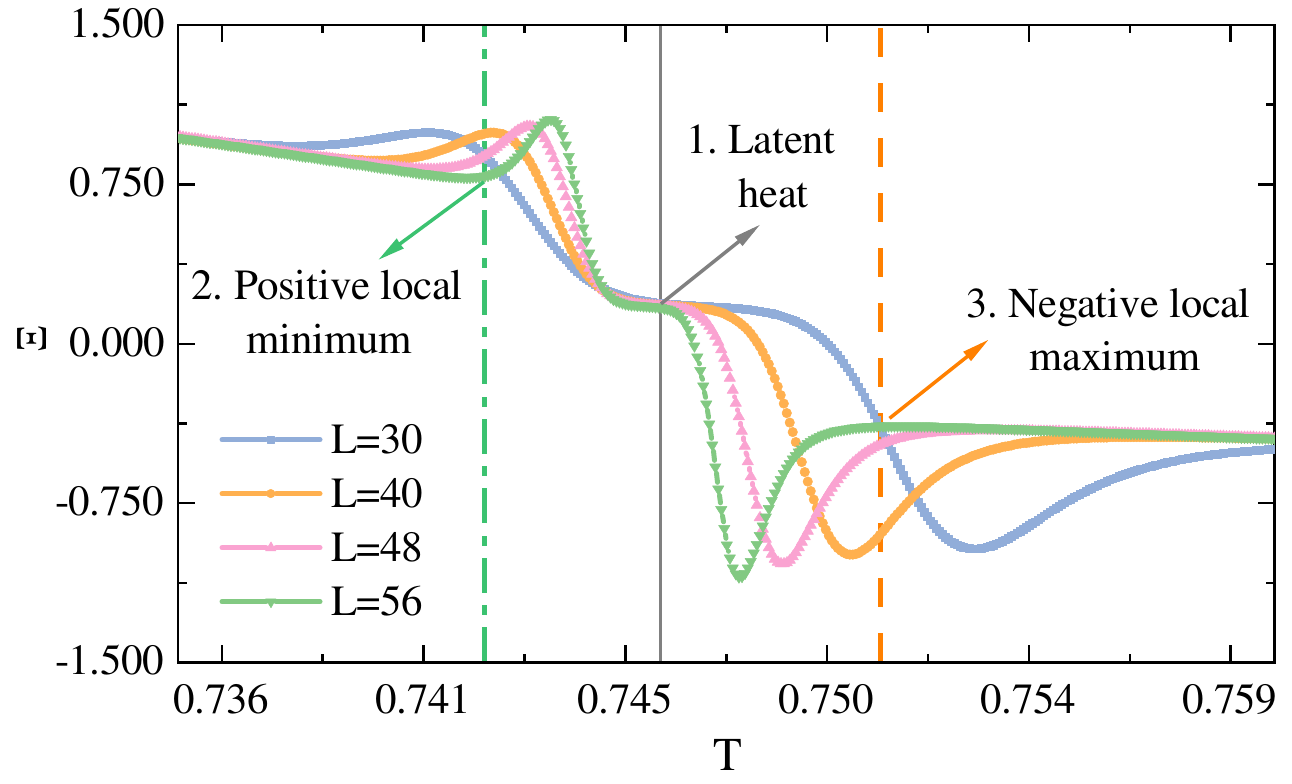}
    \caption{
    Canonical third-order diagnostic in the square-lattice eight-state Potts model.
    The primary phase transition is first order and occurs near $T_c\simeq 0.746$ for the finite-size crossing estimate. [Data were obtained from \cite{Wang2026Canonical} and replotted for comparison.]
    }
    \label{fig:potts_xi}
\end{figure}

In this regime, the canonical diagnostic $\Xi(T)$ exhibits a robust two-feature structure around the transition region, as shown in Fig.~\ref{fig:potts_xi}. Specifically, a positive local minimum appears on the low-temperature side near $T_{\mathrm{ind}}\simeq 0.742$, while a negative local maximum appears on the high-temperature side near $T_{\mathrm{dep}}\simeq 0.748$.
These two features remain stable across system sizes and are located on opposite sides of the primary transition point estimated from finite-size crossings of thermodynamic observables. The observed morphology reflects an asymmetric reshaping of the canonical energy distribution in the coexistence region.
Importantly, this two-extremum structure should be understood as a response of the fluctuation-based diagnostic in a first-order transition regime, rather than as evidence of additional thermodynamic phase transitions. The separation of the two extrema provides a convenient way to characterize the asymmetric fluctuation structure induced by phase coexistence.

This morphological behavior can also be related to standard thermodynamic signatures of first-order phase transitions. In particular, the coexistence regime is associated with latent heat and a discontinuous change in the energy landscape, which manifests in finite systems as a smoothed but rapidly varying transition between competing states. The paired-extremum structure of $\Xi(T)$ reflects this underlying redistribution process: the low-temperature extremum is associated with the onset of metastable excitation of the high-energy phase, while the high-temperature extremum corresponds to the relaxation of the low-energy phase. The region between the two extrema corresponds to the effective transition interval, within which the system undergoes a rapid but continuous change in internal energy, giving rise to a characteristic inflection-like response in fluctuation observables.

\subsection{MIPA-type third-order structures around a second-order phase transition: Ising model}
\label{subsec:ising_results}

We next consider the two-dimensional Ising model, whose primary phase transition is second order at $T_c \simeq 2.269$. In finite systems, the critical region is characterized by strong fluctuations and smooth finite-size rounding of thermodynamic quantities.

\begin{figure}
    \centering
    \includegraphics[width=0.55\linewidth]{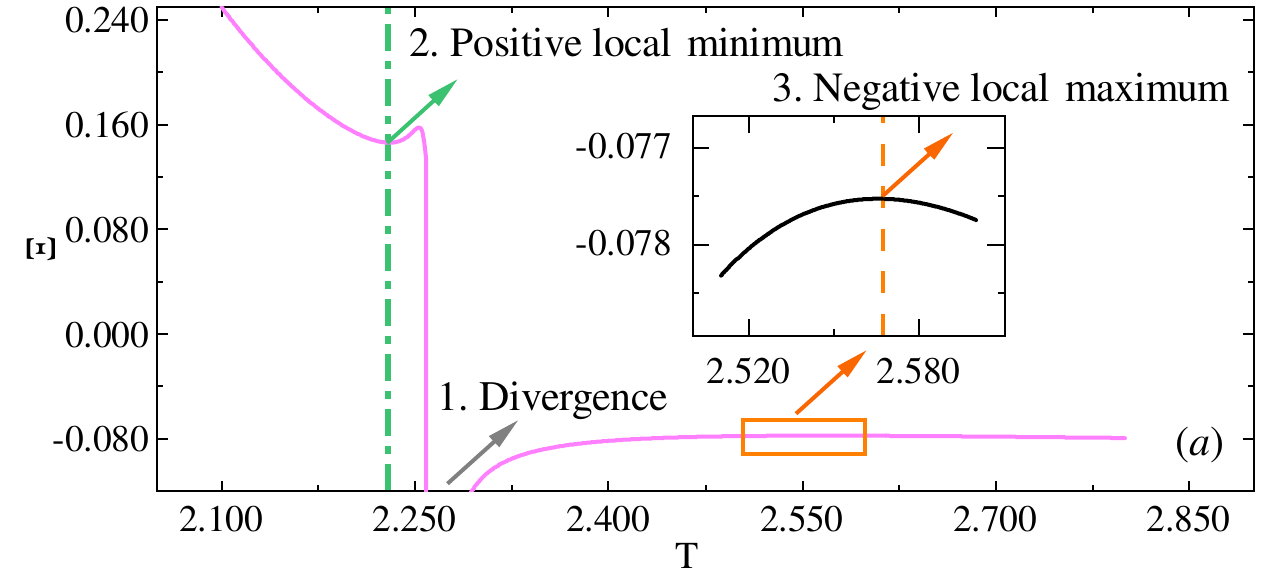}
  \caption{
Temperature dependence of the canonical fluctuation diagnostic $\Xi(T)$ in the two-dimensional Ising model. The system exhibits a second-order phase transition at $T_c \simeq 2.269$. [Data were obtained from \cite{Wang2026Canonical} and replotted for comparison.]
}
    \label{fig:ising_xi}
\end{figure}

In this regime, the canonical diagnostic $\Xi(T)$ exhibits a characteristic two-feature morphology around the critical temperature, as shown in Fig.~\ref{fig:ising_xi}. A positive local minimum appears on the low-temperature side near $T_{\mathrm{ind}}\simeq 2.229$, while a negative local maximum appears on the high-temperature side near $T_{\mathrm{dep}}\simeq 2.567$.
These two extrema are robust across system sizes and consistently appear on opposite sides of the critical point. Their locations are distinct from the finite-size estimate of $T_c$, which is obtained from standard thermodynamic observables.
The observed morphology reflects an asymmetric response of the canonical energy fluctuations in the vicinity of a second-order phase transition. In particular, the diagnostic captures changes in fluctuation asymmetry across the critical region, leading to a separation of extrema on the low- and high-temperature sides.
Importantly, this two-extremum structure should not be interpreted as indicating additional phase transitions. Instead, it represents a fluctuation-level response pattern of the canonical diagnostic in the presence of critical fluctuations.

In contrast to the first-order case, the emergence of a two-feature structure in the Ising model is not driven by phase coexistence, but by the development of long-range critical fluctuations near the continuous transition. In this regime, the divergence of the correlation length leads to a broad redistribution of fluctuation weight across temperatures, rather than a sharp separation between competing phases. As a result, the canonical diagnostic responds to the gradual restructuring of fluctuation patterns, producing a separated pair of extrema that reflects the asymmetric buildup and relaxation of critical fluctuations around the transition region. This behavior is therefore distinct from the latent-heat-driven mechanism observed in first-order transitions, and instead originates from the smooth but strongly nonlocal nature of critical correlations.

\subsection{Direct nonanalytic behavior at a third-order phase transition: ideal Bose gas}
\label{subsec:bec_jump}

We next consider ideal three-dimensional Bose--Einstein condensation, which is a third-order phase transition in the Ehrenfest classification. In the thermodynamic limit, the heat capacity remains continuous at the condensation temperature $T_c$, while its temperature derivative becomes discontinuous.

\begin{figure}
    \centering
    \includegraphics[width=0.5\linewidth]{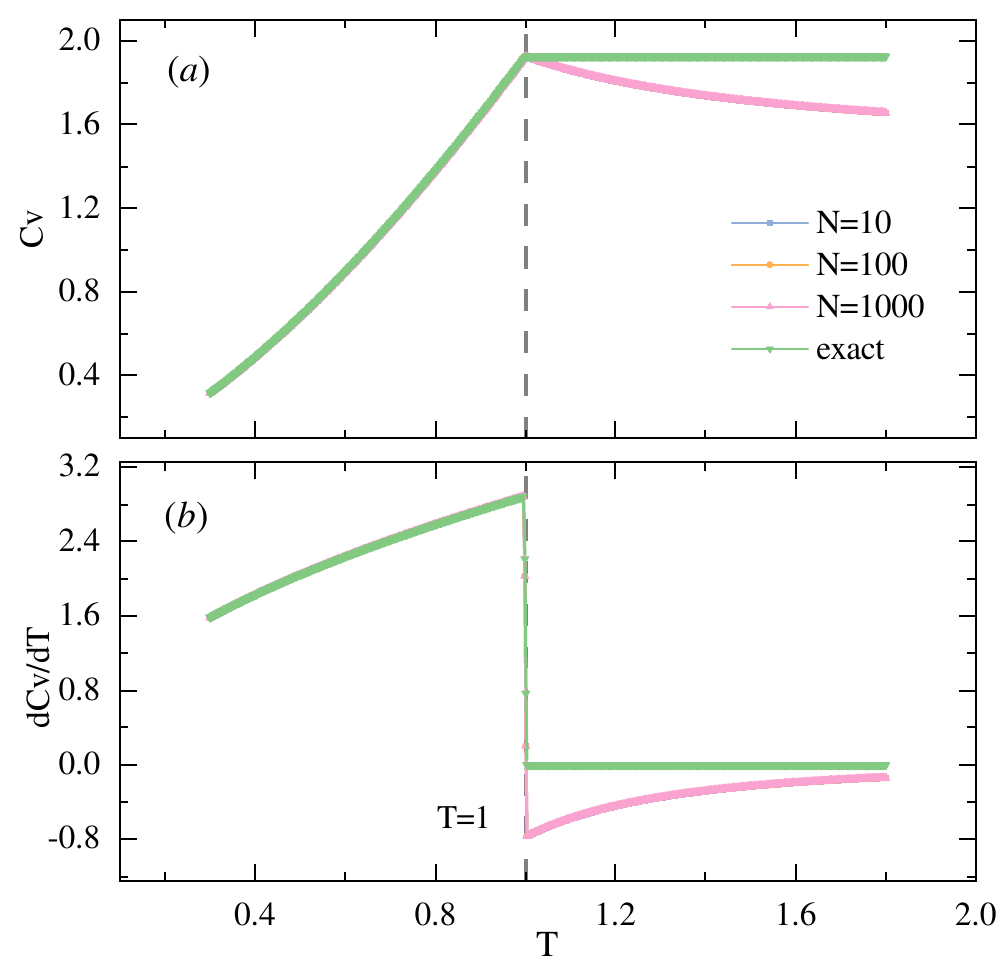}
    \caption{
    Heat-capacity signature of ideal three-dimensional Bose--Einstein condensation.
    The heat capacity $C_V$ is continuous at $T_c$, whereas its temperature derivative
    $dC_V/dT$ is discontinuous.
    This establishes the condensation transition as a third-order phase transition in
    the Ehrenfest classification, because the third temperature derivative of the free
    energy is discontinuous at $T_c$.
    }
    \label{fig:bec_cv}
\end{figure} 

As shown in Fig.~\ref{fig:bec_cv}, this implies a nonanalyticity in the third temperature derivative of the free energy. Within this framework, Bose--Einstein condensation provides a reference case in which the primary phase transition itself is third order.
The canonical cumulants are related to thermodynamic quantities through
$\kappa_2(T)=T^2 C_V(T)$ and $\kappa_3(T)=T^4 \frac{dC_V}{dT} + 2T^3 C_V(T)$.
As a result, the third-order diagnostic $\Xi(T)$ depends explicitly on both $C_V$ and its temperature derivative.

\begin{figure}
    \centering
    \includegraphics[width=0.55\linewidth]{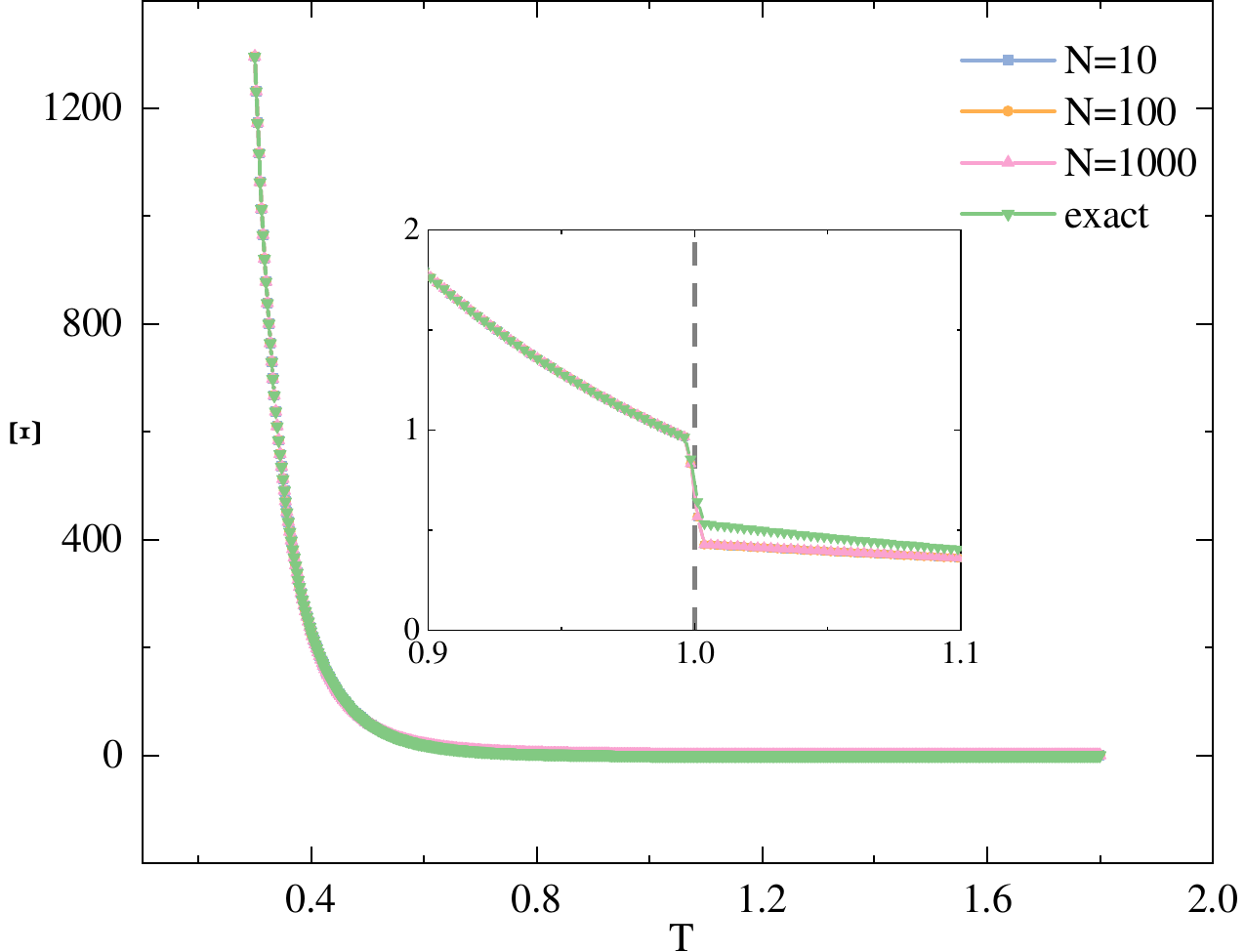}
    \caption{
    Canonical third-order diagnostic in ideal three-dimensional Bose--Einstein
    condensation.
    Since $\Xi(T)$ directly inherits the discontinuity of $dC_V/dT$ at the condensation
    temperature, the BEC morphology is a direct jump at $T_c$, not a paired-extremum
    structure.
    }
    \label{fig:bec_xi}
\end{figure}

Since $C_V$ is continuous at $T_c$ while $dC_V/dT$ is discontinuous, the behavior of $\Xi(T)$ reflects this nonanalytic structure. As shown in Fig.~\ref{fig:bec_xi}, the diagnostic exhibits a discontinuous change at $T_c$, in contrast to the smooth paired-extremum structure observed in the Ising and Potts models.
This behavior is expected because the primary phase transition itself is third order in the Ehrenfest sense. Unlike the previous two cases, where extrema arise from fluctuation-asymmetry reorganization around lower-order transitions, here the observed feature is directly associated with the thermodynamic nonanalyticity of the system in the thermodynamic limit.

This behavior can be interpreted in terms of the underlying thermodynamic mechanism of Bose--Einstein condensation. Unlike first- and second-order transitions, where the response of $\Xi(T)$ is governed by either phase coexistence or the gradual buildup of critical fluctuations, the third-order nature of the transition implies that the nonanalyticity is intrinsic to the thermodynamic free energy itself in the thermodynamic limit. As a consequence, the discontinuity in the temperature derivative of the heat capacity directly propagates into the higher-order energy cumulants, causing the canonical diagnostic to exhibit a sharp jump at the transition point rather than a pair of separated extrema. In this sense, the BEC case represents a qualitatively different response regime in which the morphology of $\Xi(T)$ reflects a direct thermodynamic singularity rather than a fluctuation-induced reorganization process.

\subsection{Morphology-based description of canonical third-order behavior}
\label{subsec:morphology_classification}

Taken together, these results reveal a clear difference in the behavior of the canonical third-order diagnostic $\Xi(T)$ across the three representative systems. Rather than focusing on classification of phase transitions, we summarize the observed behavior in terms of the morphology of the fluctuation-based observable.
For the two-dimensional Ising model and the eight-state Potts model, $\Xi(T)$ exhibits a characteristic structure consisting of a positive local minimum on the low-temperature side and a negative local maximum on the high-temperature side of the primary transition region. This paired-extremum structure is robust across system sizes and appears in both second-order and first-order phase transition regimes.

In contrast, for ideal three-dimensional Bose--Einstein condensation, $\Xi(T)$ displays a discontinuous change at the condensation temperature in the thermodynamic limit. This behavior reflects the nonanalyticity of the third derivative of the free energy, rather than fluctuation asymmetry around a lower-order transition.
These results indicate that $\Xi(T)$ exhibits distinct response morphologies across different thermodynamic regimes. 
In the representative systems considered here, first- and second-order phase transitions are associated with paired-extremum responses of $\Xi(T)$, whereas the third-order Bose--Einstein condensation transition shows a discontinuous response.

\begin{table}
    \centering
   \caption{
Summary of the response morphologies of $\Xi(T)$ in the representative systems considered here.
}
    \label{tab:morphology_classification}

    \small
    \setlength{\tabcolsep}{4pt}

    \begin{tabularx}{\linewidth}{
        >{\raggedright\arraybackslash}p{0.21\linewidth}
        >{\raggedright\arraybackslash}p{0.20\linewidth}
        >{\raggedright\arraybackslash}p{0.22\linewidth}
        >{\raggedright\arraybackslash}X
    }
        \toprule
        Phase transition 
        & System 
        & Morphology of $\Xi(T)$ 
        & Interpretation \\
        \midrule

                First-order phase transition
        & 8-state Potts model
        & Paired extrema
        & Fluctuation redistribution induced by phase coexistence \\

        Second-order phase transition
        & 2D Ising model
        & Paired extrema
        & Fluctuation redistribution driven by critical correlations \\

        Third-order phase transition
        & Ideal 3D Bose gas
        & Direct jump
        & Response governed by thermodynamic nonanalyticity \\

        \bottomrule
    \end{tabularx}
\end{table}

\section{Discussion}
\label{sec:discussion}

The present comparison suggests that the morphology of $\Xi(T)$ is strongly influenced by the mechanism through which thermodynamic fluctuations are reorganized.
In the eight-state Potts model, the paired-extremum response is associated with phase coexistence and latent-heat-driven redistribution of the energy distribution. 
In the two-dimensional Ising model, a similar paired-extremum morphology arises from the buildup and relaxation of critical fluctuations, rather than from coexistence. 
Thus, the same qualitative morphology may appear in different lower-order transition regimes, while reflecting distinct microscopic mechanisms.

The ideal Bose gas provides a contrasting case. 
Because the primary nonanalyticity itself occurs in the third temperature derivative of the free energy, the discontinuity of $dC_V/dT$ is directly inherited by the cumulant ratio $\Xi(T)$. 
The resulting response is therefore a direct jump rather than a separated pair of extrema. 
This distinction suggests that $\Xi(T)$ should not be used as a simple classifier of Ehrenfest order. 
Instead, it is better understood as a fluctuation-based diagnostic whose morphology reveals how energy fluctuations are reorganized near a thermodynamic transition.

\section{Conclusion}
\label{sec:conclusion}
We have compared the response morphology of a fixed canonical fluctuation diagnostic, $\Xi(T)$, across representative first-, second-, and third-order phase transitions. 
For the eight-state Potts model and the two-dimensional Ising model, $\Xi(T)$ exhibits a paired-extremum structure around the primary transition region, reflecting fluctuation redistribution associated with phase coexistence or critical correlations. 
For ideal three-dimensional Bose--Einstein condensation, by contrast, $\Xi(T)$ shows a direct jump at the condensation temperature, because the primary thermodynamic nonanalyticity itself is third order.

These results indicate that $\Xi(T)$ does not provide a direct classification of Ehrenfest order. 
Rather, its morphology offers a unified fluctuation-based perspective for comparing how different thermodynamic mechanisms shape higher-order energy-fluctuation responses.

\bibliographystyle{unsrtnat}
\bibliography{references}

\end{document}